\documentclass[12pt,a4paper]{article}

\usepackage[utf8]{inputenc}
\usepackage[T1]{fontenc}
\usepackage[margin=1in]{geometry}
\usepackage{amsmath,amssymb,bm}
\usepackage{newtxtext,newtxmath}
\usepackage{graphicx}
\usepackage{setspace}
\usepackage[font=small]{caption}
\usepackage{ragged2e}
\usepackage{xcolor}
\usepackage{cite}

\graphicspath{{./}{figures/}}

\justifying

\newcommand{\bk}{\bm{k}}
\newcommand{\bn}{\bm{n}}
\newcommand{\bmagn}{\bm{m}}

\begin{document}



\begin{center}
{\LARGE \bfseries  Spin superfluid transport of magnons in altermagnetic insulators}

\vspace{1.2em}
{\large Dele Li$^{1}$, Meng-Han Zhang$^{1}$, Dao-Xin Yao$^{1,\dagger}$}

\vspace{0.6em}
{\normalsize $^{1}$Guangdong Provincial Key Laboratory of Magnetoelectric Physics and Devices, State Key Laboratory of Optoelectronic Materials and Technologies, Institute of Neutron Science and Technology, School of Physics, Sun Yat-Sen University, Guangzhou, 510275, China \\
$^{1}$These authors contributed equally to this work.\\
$^{\dagger}$Corresponding author: yaodaox@mail.sysu.edu.cn}
\end{center}

\vspace{1em}

\begin{abstract}
\noindent
We study the spin superfluidity for Lieb and inverse Lieb lattices in altermagnets, 
which enables coherent transport of spin angular momentum
carried by the phase of an easy-plane N\'eel order. 
The easy-plane anisotropy supplies the $U(1)$ platform, 
while the nonrelativistic spin splitting generates 
sublattice-alternating exchange imbalance with $B_{1g}$ symmetry. 
Via a continuum theory and Landau--Lifshitz--Gilbert simulations, 
we find that the Goldstone mode becomes quadratic 
at the critical stability boundary. 
An axial phase gradient transfers phase-channel spectral weight to the gapped mode,
whereas a diagonal flow leaves these signatures absent.
The spin superfluid exhibits momentum-space anisotropy 
due to the chiral-splitting magnon dispersion, while strong 
fields or elevated temperatures suppress phase coherence.
Accounting for the contribution of thermal magnons at the interface, 
we present that the critical spin flow varies with the injection angle 
under the altermagnetic crystal-field effects.
The spin-superfluid decay length scales with the interfacial spin-mixing conductance, 
which can be measured by the spin Hall magnetoresistance and spin Seebeck effect.
Our findings establish altermagnetic insulators, 
particularly $\mathrm{La_2O_3FeMnSe_2}$, 
as promising platforms for low-dissipation spin transport.
\end{abstract}

\vspace{0.5em}

\section*{Introduction}

Spin superfluidity, the dissipationless transport of spin angular momentum
driven by a phase gradient of an easy-plane magnetic order parameter~\cite{Sonin2010,Takei2014FM,Takei2014}, is the
magnetic analogue of superfluidity in helium and of superconductivity in
electronic systems~\cite{NewRef07,NewRef12,NewRef13,NewRef14,NewRef15}. Realising it requires
long-range coherent transport of spin angular momentum, carried by the
spatially varying phase of the order parameter~\cite{Sonin2019,Evers2020,Chen2014,Takei2015,Li2021,NewRef01,NewRef08,Goli2021}. Altermagnets are a recently
identified class of magnetic phases with no net magnetisation yet a
nonrelativistic spin splitting in reciprocal
space~\cite{Smejkal2022,NewRef02,McClarty2024,NewRef03,NewRef04,NewRef01,Lundemo2025,Liu2022SpinGroup,Chen2024SpinSpace,Liu2026Oriented}. Unlike conventional collinear
antiferromagnets, their momentum-dependent spin splitting is protected by
crystal rotation or mirror symmetry~\cite{Smejkal2022,McClarty2024,NewRef04,Chen2025Magnons,Zhu2025Geometry}, so that spin currents can be generated without an
external magnetic field and new correlated states become accessible. The
oxychalcogenide $\mathrm{La_2O_3FeMnSe_2}$ is an altermagnetic
insulator with unconventional magnon excitations~\cite{Chen2025Magnons}.
Here we consider Lieb and inverse Lieb lattice descriptions and a $d_{x^2-y^2}$-wave-like altermagnetic splitting. It is therefore
natural to ask how a nonrelativistic, momentum-selective altermagnetic
splitting modifies the collective dynamics of a spin superfluid, a question
not addressed by conventional antiferromagnets. For the exchange
pattern considered here, the relevant long-wavelength form factor transforms
as $B_{1g}$ and scales as $k_x^2-k_y^2$. Combining
analytical modelling with Landau--Lifshitz--Gilbert (LLG) simulations, we
identify the universal signatures and the altermagnetic fingerprints of the
collective spin dynamics of the spin superfluid.

Interfacial physics has attracted increasing attention, 
particularly for exploiting the non-Hermitian pumping and dissipation.
While the influence of bulk damping on superfluid spin transport has been
examined~\cite{Flebus2016,Brataas2020,Yuan2018Cr2O3,Yuan2026CrCl3,Leiviska2025}, 
the role of boundary conditions remains largely unexplored. 
Here we investigate the spin-superfluid state in a hybrid normal--metal altermagnetic insulator heterostructure, 
focusing on interfacial electrical and thermal effects to drive collective spin dynamics~\cite{Moura2026,Fischer2018,Qiu2013,NewRef09,NewRef11,NewRef16}.
A Zeeman field reduces the relaxation time and thereby accelerates the approach to the steady state, 
whereas both the spin and phase stiffnesses decrease with rising temperature. 
Incorporating interfacial STT into the LLG simulations, we reveal the 
microscopic origin of the boundary conditions and bulk state 
with the altermagnetic order parameter~\cite{Manchon2019,Mandokoro2025,NewRef10}.
The spin-mixing and spin Seebeck conductances jointly determine 
the accumulated spin angular momentum, 
thereby generating a torque on the N\'{e}el vector, 
a prerequisite for spintronic applications\cite{Zhu2025SQUID,Schwartz2022}.


\section*{Results}

\subsection*{Continuum model of the altermagnetic spin superfluid}

\begin{figure}[htbp]
 \centering
 \includegraphics[width=1 \textwidth]{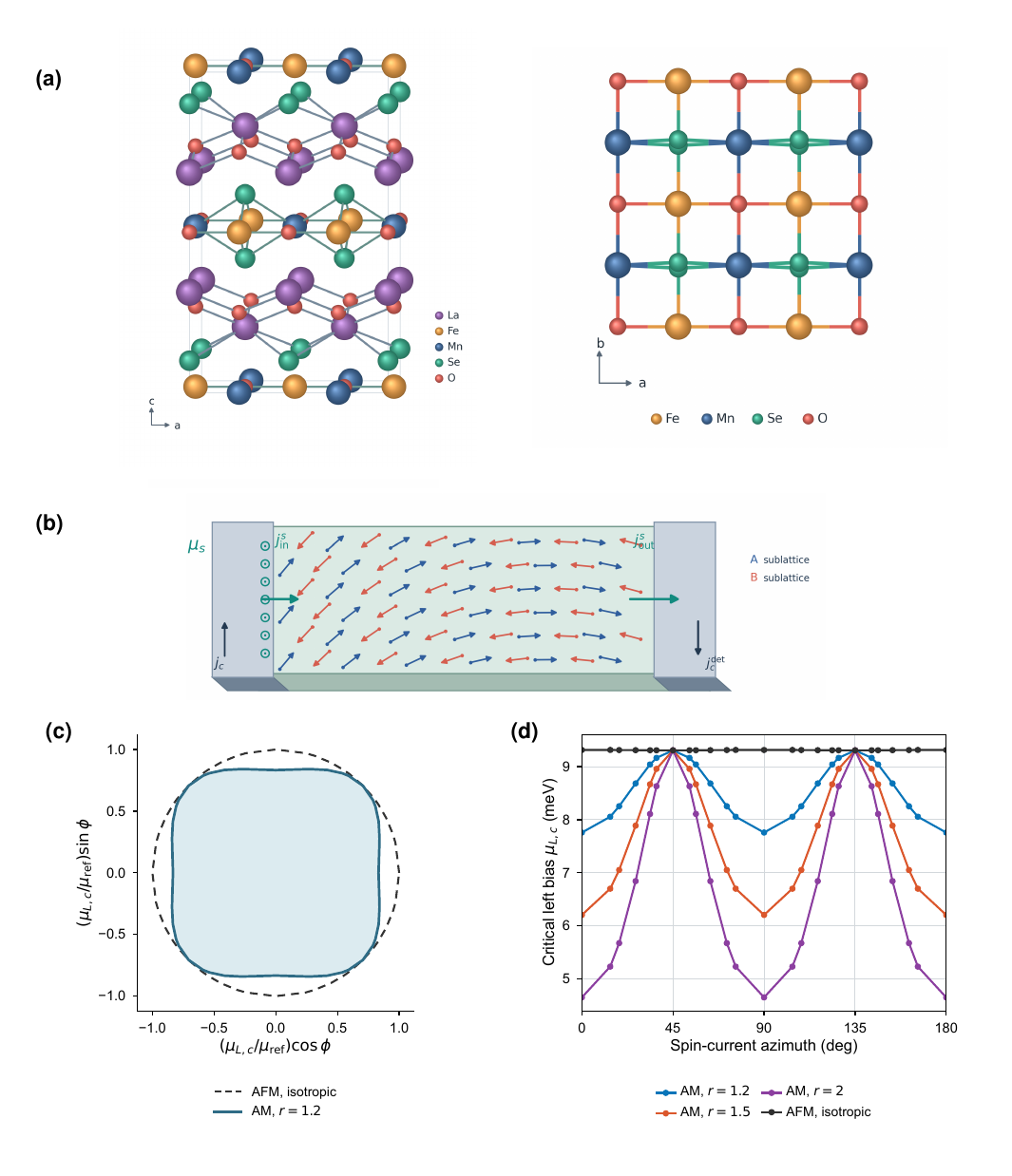}
 \caption{Altermagnetic spin-superfluid and the directional dependence of the critical spin bias. 
 \textbf{(a)} Crystal structure and lattice representation of the altermagnetic insulator $\mathrm{La_2O_3FeMnSe_2}$. 
 \textbf{(b)} Schematic of the metal--altermagnet--metal heterostructure. 
 A spin accumulation $\mu_s$ at the left interface injects spin angular momentum into the altermagnetic layer. 
 Blue and red arrows indicate the spin orientations on the two magnetic sublattices. 
 \textbf{(c)} Polar representation of the normalized left spin bias 
 $\mu_{L,c}/\mu_{\mathrm{ref}}$ required to reach the critical spin flow, 
 as a function of the spin-current azimuth $\phi$. 
 The solid curve corresponds to the altermagnetic case with $r=1.2$, 
 and the dashed circle denotes the isotropic antiferromagnetic reference. 
 \textbf{(d)} Critical left spin bias $\mu_{L,c}$ as a function of 
 the spin-current azimuth for different values of $r$; 
 the black curve shows the isotropic antiferromagnetic reference. 
 To enhance the altermagnetic response, 
 a $d$-wave-type angular dependence of the left-contact damping is adopted, 
 where $r$ is the ratio of the axial to diagonal contact damping.}
 \label{fig:structure}
\end{figure}

Motivated by the unconventional chiral magnon splitting in $\mathrm{La_2O_3FeMnSe_2}$ 
(Fig.\ref{fig:structure}(a)), 
we consider a two-dimensional $d$-wave altermagnetic model with two collinear magnetic sublattices.
The Lieb and inverse Lieb lattices are constructed by the spin space group (SSG)
$I^{-1}4/^{1}m^{1}m^{-1}m^{\infty m}1$ ( nSSG:139.71.1.1.L ), 
in which the two sublattices are related by the combined symmetry $\mathcal{M}\tau$, 
where $\mathcal{M}$ denotes a mirror operation and $\tau$ a half-unit-cell translation. 
The Hamiltonian governing the spin-superfluid dynamics is

\begin{align}
 \mathcal H_0={}&J_1\sum_{\langle ij\rangle}\bm S_i\cdot\bm S_j
 +\sum_{\mu=m, n}\!J_{2\mu}\sum_{\langle\!\langle ij\rangle\!\rangle}
 \bm S_i\cdot\bm S_j \nonumber+K_\perp\sum_i(S_i^z)^2+h_B\sum_i S_i^z,
\end{align}

\begin{figure}[htbp]
 \centering
 \includegraphics[width=1 \textwidth]{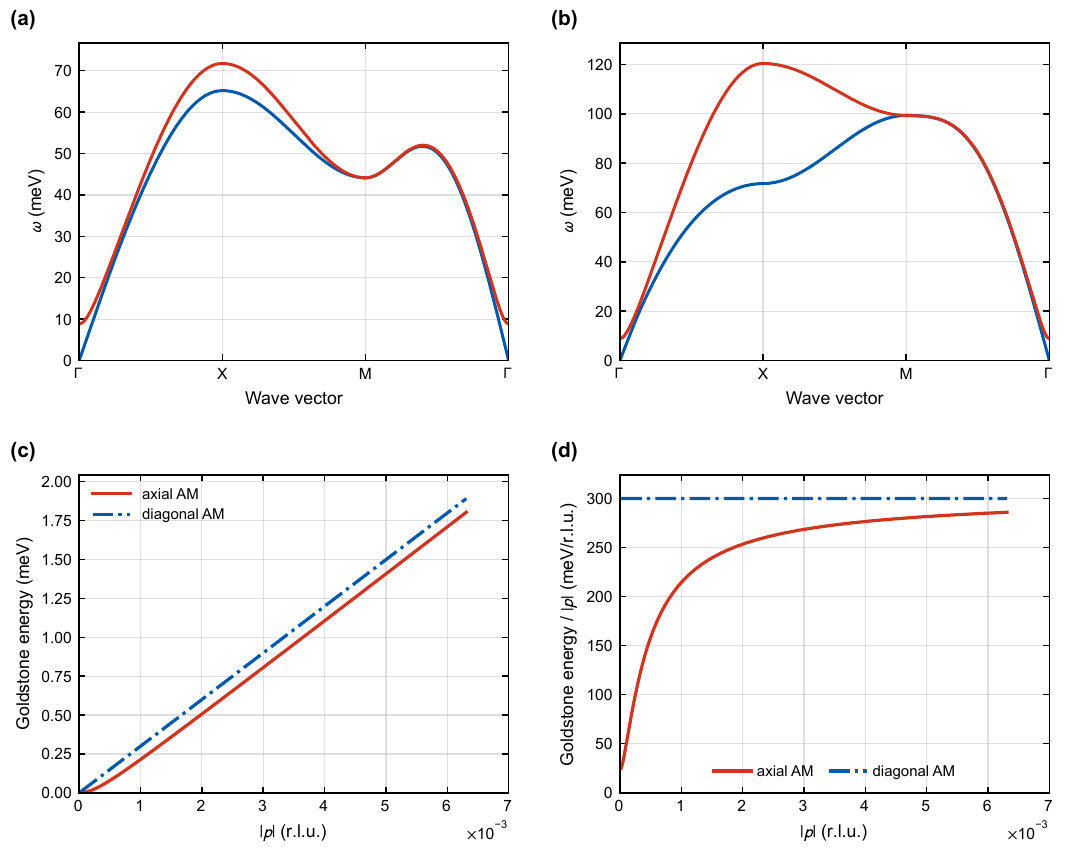}
 \caption{Magnon dispersions and directional dependence of the Goldstone mode at the critical spin flow. \textbf{(a), (b)} Magnon dispersions along the high-symmetry path $\Gamma$--X--M--$\Gamma$ for the Lieb lattice \textbf{(a)} and the inverse Lieb lattice \textbf{(b)}. The red and blue curves represent the two magnon branches. \textbf{(c)} Low-energy Goldstone-mode dispersion for axial and diagonal spin superflow, with each flow set to its respective critical value. The axial-flow mode becomes quadratic in the long-wavelength limit, while the diagonal-flow mode remains linear. \textbf{(d)} Goldstone-mode energy divided by the wave-vector magnitude, $\omega_{\mathrm G}/|p|$, under the same critical-flow conditions, where $p$ denotes the probe wave-vector magnitude $k$. In the long-wavelength limit, this ratio vanishes for axial flow and remains finite for diagonal flow.}
 \label{fig:magnons}
\end{figure}

where $J_1>0$ is the nearest-neighbour antiferromagnetic coupling. 
Ferromagnetic next-nearest-neighbor bonds connect sites on 
the same magnetic sublattice, 
with $J_{Ai,Aj}^{ix,jx}=J_{Bi,Bj}^{iy,jy}=J_{2m}$ and $J_{Ai,Aj}^{iy,jy}=J_{Bi,Bj}^{ix,jx}=J_{2n}$.
DFT calculations yield $J_1=9.261$ meV, $J_{2m}=2.107$ meV, 
$J_{2n}=2.762$ meV, and $K_\perp=0.0852$ meV. 
The resulting magnon band structures for the Lieb and inverse-Lieb lattices 
are presented in Figs.~\ref{fig:magnons}(a) and \ref{fig:magnons}(b), respectively.
The positive single-ion term $K_\perp>0$ favours easy-plane magnetization, 
providing a realisation of the spin-superfluid state.
Consequently, the in-plane phase of the low-energy N\'eel field denotes a Goldstone mode 
canonically conjugate to the uniform out-of-plane spin density, 
whereas out-of-plane Néel fluctuations form a distinct gapped collective mode. 

We formulate a continuum theory for this microscopic model, 
introducing the N\'eel and uniform-magnetisation fields by
$\frac{\bm{S}_A}{S}=\bn+\bmagn$ and $\frac{\bm{S}_B}{S}=-\bn+\bmagn$.
The two magnetic sublattices are interchanged 
by a fourfold rotation $C_4$ combined with spin reversal, 
whereas the inequivalent exchanges $J_{2m}$ and $J_{2n}$ 
generate an altermagnetic continuum coefficient 
$\lambda\propto (J_{2m}-J_{2n})$.
With $D\equiv\partial_x^2-\partial_y^2$, 
constrained elimination of the uniform magnetization 
produces a $B_{1g}$ dynamical–gradient term: 
easy-plane anisotropy underpins the $U(1)$ spin superfluidity 
and out-of-plane gap, while the $B_{1g}$ form factor arises 
purely from the nonrelativistic altermagnetic coupling. 
 Variation subject to $\bn\cdot\bmagn=0$ yields
$ P_\perp=\mathbf{1}-\bn\bn^{\mathsf T}$, 
and the effective Lagrangian is then

\begin{align}
 \mathcal L_{\rm eff}={}&\frac{\rho}{2}(\partial_t\bn)^2
 +g_{\rm AM}(\bn\times\partial_t\bn)\cdot D\bn
 +\frac{g_{\rm AM}^2}{2\rho}\left|P_\perp D\bn\right|^2 \nonumber
 -\frac{A_{\rm ex}}{2}(\nabla\bn)^2-\frac{K}{2}n_z^2,
\end{align}

Here $\rho=\chi s^2$ is the inertial coefficient and $g_{\rm AM}=\chi s\lambda$ is the altermagnetic dynamical coupling coefficient.
The parameters $s$, $\chi$, and $A_{\rm ex}$ denote the Berry-phase coefficient, the transverse susceptibility of the uniform magnetisation, and the exchange stiffness, respectively. The coefficient $K$ denotes the out-of-plane anisotropy in the continuum theory, while $K_\perp$ denotes the single-ion anisotropy in the lattice model.
The spin-injection and transport geometry considered here is shown in Fig.~\ref{fig:structure}(b); 
further derivations are provided in Methods.

\subsection*{Landau--Lifshitz--Gilbert dynamics}

A quantitative comparison with a complete lattice spin-wave calculation is
beyond the minimal continuum treatment presented here. We therefore
incorporate Landau--Lifshitz--Gilbert (LLG) phenomenology to model
numerically the N\'eel-vector dynamics of this altermagnet insulator. 
For either sublattice,

\begin{align}
 \dot{\bm S}_{A,\bm r}
 &=-\gamma\bm S_{A,\bm r}\times\bm B^{\mathrm{eff}}_{A,\bm r}
 +\alpha_{\bm r}\bm S_{A,\bm r}\times\dot{\bm S}_{A,\bm r}, \nonumber\\
 \dot{\bm S}_{B,\bm r}
 &=-\gamma\bm S_{B,\bm r}\times\bm B^{\mathrm{eff}}_{B,\bm r}
 +\alpha_{\bm r}\bm S_{B,\bm r}\times\dot{\bm S}_{B,\bm r}.
 \label{eq:LLG}
\end{align}

Here, $\gamma=g\mu_B/\hbar$ is the gyromagnetic ratio, 
$\alpha_{\bm r}$ is the local Gilbert damping and 
$\bm B^{\mathrm{eff}}_{\nu,\bm r}=\bm h_{\nu,\bm r}/(g\mu_B S)$ is 
the effective field acting on the two sublattices $\nu=A,B$. 
We resolve the spin-transfer torque (STT) contributions arising from 
thermal magnons and from the interfacial spin-mixing conductance, 
which is accessible via spin Hall magnetoresistance (SMR).
The directional dependence of the critical left spin bias is shown in Figs.~\ref{fig:structure}(c) and \ref{fig:structure}(d).
To highlight the altermagnetic response, we adopt a $d$-wave-type angular dependence of the left-contact damping, with different ratios of axial to diagonal damping. 
The two sublattices carry opposite spin angular momenta, 
$s_A(k)\approx+\hbar$ and $s_B(k)\approx-\hbar$, 
and experience opposite torques from the effective field $\bm h_{\nu,\bm r}$.

\begin{align}
 \tau_{A,\bm r}&=\gamma B_{\rm FL}\, \bm S_{A,\bm r} \times \mathbf p_{\rm s}
 + \gamma B_{\rm DL}\, \bm S_{A,\bm r} \times (\mathbf p_{\rm s} \times \bm S_{A,\bm r}), \nonumber\\
 \tau_{B,\bm r}&=\gamma B_{\rm FL}\, \bm S_{B,\bm r} \times \mathbf p_{\rm s}
 + \gamma B_{\rm DL}\, \bm S_{B,\bm r} \times (\mathbf p_{\rm s} \times \bm S_{B,\bm r}),
 \label{eq:STT}
\end{align}

where $\mathbf p_{\rm s}$ is the spin-polarisation axis, 
with field-like and damping-like amplitudes $B_{\rm FL}$ and $B_{\rm DL}$. 
The altermagnetic ordering also influences the interfacial spin-mixing conductance, 
which enters superfluid spin transport as a non-Hermitian injection term. 
In metal--altermagnet heterostructures, the magnon-related contributions 
are governed by the deviation of the magnon chemical potential $\mu_s$ 
in the vicinity of the interfaces.

\begin{equation}
 \mu_s \simeq \sum_\eta \int d\omega\, D_\eta(\omega)\, s_{\eta}(\omega)\,
 \delta n_\eta(\omega),
 \label{eq:Chemical-potential}
\end{equation}

where $\eta$ labels magnon branches and $D_\eta(\omega)$ is the magnon density of states, 
and $\delta n_\eta(\omega_\eta)= n_\eta(\omega_\eta) - n_B\!\left(\omega_\eta\right)$ 
can be expanded by the temperature gradient.
Thermal spin currents inherit a momentum-space anisotropy 
from the spin Seebeck effect, and
ultimately convert into a coherent spin superfluid. 
This superfluid decays algebraically with 
a characteristic length $L \propto \chi_{\bm r}/\alpha_{\bm r}$, 
while $\delta n_\eta(\bm k)$ is described by the Boltzmann equation

\begin{figure}[htbp]
 \centering
 \includegraphics[width=1 \textwidth]{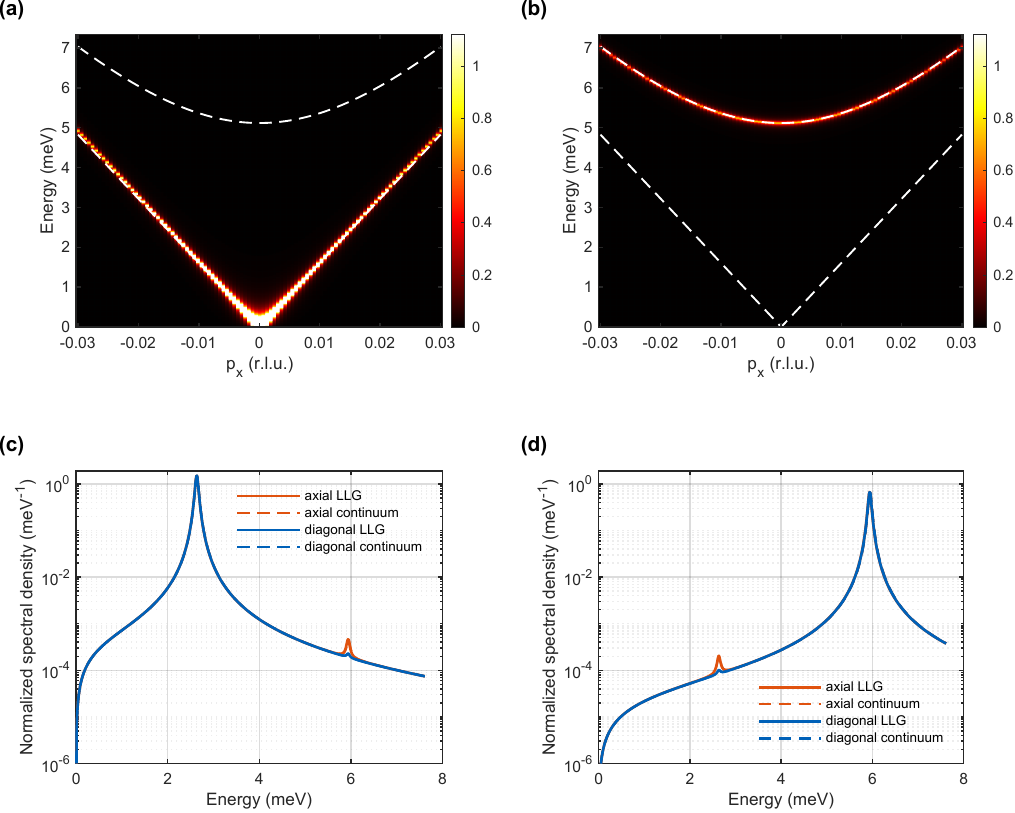}
 \caption{Spectral response in the phase and out-of-plane channels. \textbf{(a), (b)} Energy- and momentum-resolved spectral intensity of the antiferromagnet in the phase channel \textbf{(a)} and out-of-plane channel \textbf{(b)} along the $p_x$ direction, where $p_x$ denotes the corresponding probe wave-vector component $k_x$. The colour scale indicates the spectral intensity, and the white dashed curves show the dispersions of the two modes. \textbf{(c), (d)} Normalized spectral density of the altermagnet as a function of energy in the phase channel \textbf{(c)} and out-of-plane channel \textbf{(d)}. Orange and blue curves correspond to axial and diagonal spin superflow, respectively. Solid and dashed curves denote the results from Landau--Lifshitz--Gilbert (LLG) simulations and the continuum theory, respectively.}
 \label{fig:weights}
\end{figure}
\begin{figure}[htbp]
 \centering
 \includegraphics[width=1.0\textwidth]{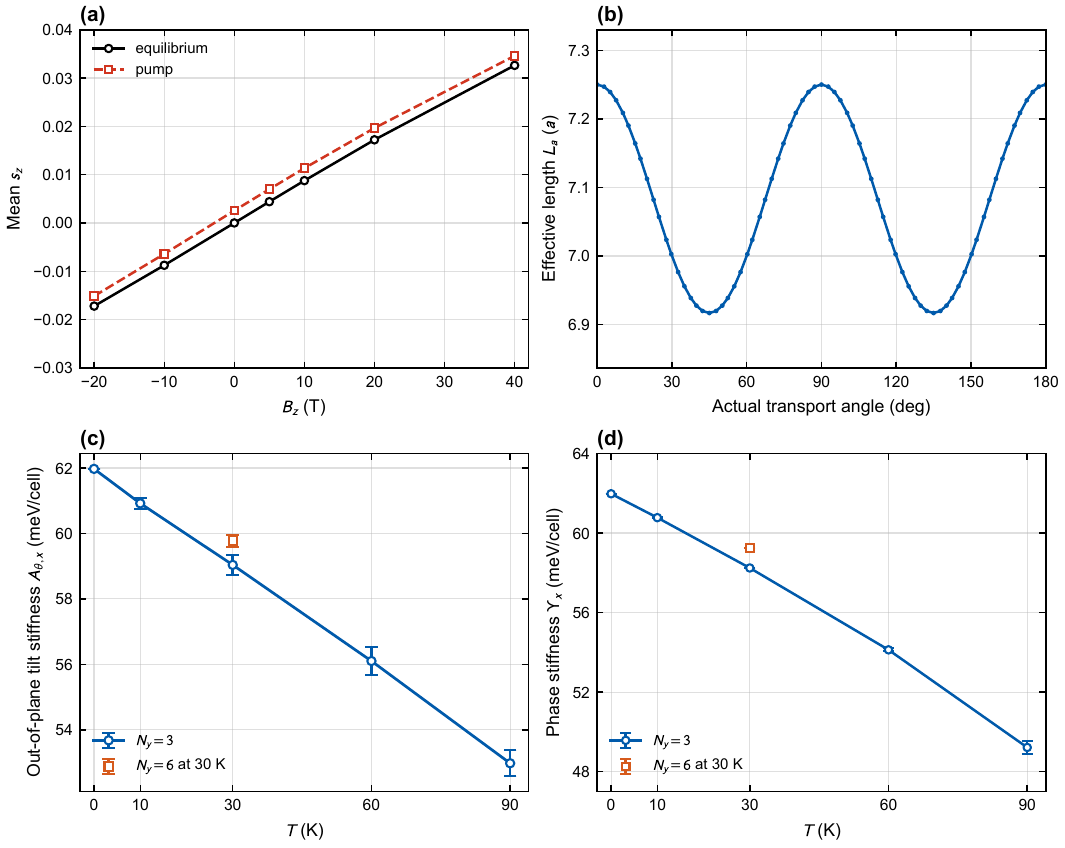}
 \caption{Magnetic-field, directional and temperature dependence of spin-superfluid transport. \textbf{(a)} Mean out-of-plane spin component $s_z$ as a function of the magnetic field $B_z$, comparing equilibrium with the case of interfacial spin injection. \textbf{(b)} Characteristic decay length $L_a$, in units of the lattice constant $a$, as a function of the transport direction. \textbf{(c)} Temperature dependence of the spin stiffness. \textbf{(d)} Temperature dependence of the phase stiffness for $N_y=3$, together with a result for $N_y=6$ at $T=30$ K.}
 \label{fig:fieldtemp}
\end{figure}


\begin{equation}
 \bm v_\eta(\bm k) \cdot \nabla n_\eta(\bm k) = \left( \frac{\partial n_\eta}{\partial t} \right)_{\text{scatt}}
 + \text{injection term},
 \label{eq:Boltzmann-equation}
\end{equation}

where the group velocity $\bm v_\eta(\bm k)=\nabla_{\bm k}\omega_\eta(\bm k)$ 
follows directly from the dispersion. 
The scattering rate and density of states are also governed by 
the nonrelativistic chiral splitting of the magnon bands. 
Fig.~\ref{fig:fieldtemp} shows that a Zeeman field shortens the relaxation time 
and accelerates equilibration toward the spin-superfluid steady state, 
whereas increasing temperature reduces both the spin and phase stiffness 
and eventually suppresses superfluidity.


\subsection*{Goldstone velocity and spectral-weight transfer}

Altermagnetic coupling hybridises the phase and out-of-plane fluctuations. 
In the spin-superfluid state, the two collective modes admit the long-wavelength expansions:

\begin{align}
 \omega_-^2(\bk)&=\frac{A_{\rm ex}\widetilde M}{\rho M}k^2+O(k^4),&
 \omega_+^2(\bk)&=\frac{M}{\rho}
 +\left(\frac{A_{\rm ex}}{\rho}+\frac{A_{\rm ex}g_{\rm AM}^2Q^2}{\rho^2M}\right)k^2
 -\frac{2g_{\rm AM}^2Q}{\rho^2}\delta_{\bk}+O(k^4).
 \label{eq:dispersions}
\end{align}

Here, $\bm q$ is the background phase gradient and $q=|\bm q|$.
We define $Q\equiv q_x^2-q_y^2$, $M\equiv K-A_{\rm ex}q^2$, and
$\widetilde M\equiv M-g_{\rm AM}^2Q^2/\rho$. The quantity $\delta_{\bk}\equiv k^2\cos(2\phi_{\bk})$ 
is the leading $B_{1g}$ form factor of the probe wave vector $\bk$, 
with $k=|\bk|$ and $\phi_{\bk}$ its in-plane angle. The gapped-mode curvature also 
acquires a direction-dependent term $\propto Q\delta_{\bk}$. 
Both modes therefore disperse anisotropically with the superflow direction (Eq.~\eqref{eq:dispersions}). 
Axial and diagonal flows represent two limiting cases. 
For axial flow, $|Q|$ is maximal at fixed $|\bm q|$, 
and the altermagnetic coupling suppresses the Goldstone velocity. 
For diagonal flow, $Q=0$ and the Goldstone velocity is unchanged at this order, 
as for a conventional antiferromagnet.

Long-wavelength stability requires $\widetilde M>0$ for axial flow 
and $M>0$ for diagonal flow; 
axial flow therefore reaches its critical gradient first. 
As $\widetilde M\to0^+$ at the axial boundary, the Goldstone velocity vanishes, 
whereas the upper branch remains gapped at $\bk=\bm{0}$, with $\omega_+^2(\bm{0})=M/\rho$. 
The Goldstone mode then acquires a quadratic dispersion:

\begin{equation}
 \omega_-^2(\bk)\simeq
 \frac{A_{\rm ex} k^2}{g_{\rm AM}^2Q^2}
 \left(A_{\rm ex}k^2+\frac{2g_{\rm AM}^2Q}{\rho}\delta_{\bk}\right)+O(k^6).
 \label{eq:critical-mode}
\end{equation}

By contrast, diagonal flow ($Q=0$) and the conventional antiferromagnetic limit ($g_{\rm AM}=0$) 
have $M\to0^+$ as the stability boundary: the Goldstone branch remains linear to 
leading order, whereas the out-of-plane-polarised upper branch softens at $\bk=\bm{0}$. 
Figure. 2(c) and 2(d) contrasts the quadratic axial critical mode with its linear diagonal counterpart. 
Because the coefficient in Eq.~\eqref{eq:critical-mode} depends on the propagation direction, 
stability at finite wave vector requires checking the full spectrum within the gradient expansion.

The altermagnetic coupling also redistributes spectral weight between 
the two modes within each response channel. Introducing the phase fluctuation $\varphi$ and out-of-plane fluctuation $\theta$, and their sources 
$f_\varphi\varphi+f_\theta\theta$ yields retarded susceptibilities 
in the in-plane phase ($\varphi$) and out-of-plane ($\theta$) channels, 
each with two poles. Their Kubo residues $\mathcal R_\nu^a$ ($\nu=\pm$, 
$a=\varphi,\theta$) obey $\mathcal R_-^a+\mathcal R_+^a=1/\rho$, 
confirming a redistribution of weight between existing modes rather than 
the emergence of new ones. For axial flow, the residues read at long wavelength:

\begin{align}
 \mathcal R_-^\theta(\bk)&=
 \frac{A_{\rm ex}g_{\rm AM}^2Q^2}{\rho^2M^2}k^2+O(k^4),\qquad
 \mathcal R_+^\theta(\bk)= \frac{1}{\rho}-\frac{A_{\rm ex}g_{\rm AM}^2Q^2}{\rho^2M^2}k^2+O(k^4),
 \label{eq:Qnonzero-long-wavelength-theta}\\
 \mathcal R_-^\varphi(\bk)&=
 \frac{\widetilde M}{\rho M}+O(k^2),\qquad
 \mathcal R_+^\varphi(\bk)= \frac{g_{\rm AM}^2Q^2}{\rho^2M}+O(k^2).
 \label{eq:Qnonzero-long-wavelength-phase}
\end{align}

Axial flow transfers a fraction $g_{\rm AM}^2Q^2/(\rho M)$ of the phase-channel weight 
from the Goldstone mode to the gapped mode, endowing the latter with 
a finite residue at zero probe momentum (Eqs.\eqref{eq:Qnonzero-long-wavelength-theta} and\eqref{eq:Qnonzero-long-wavelength-phase}). 
In the out-of-plane channel, by contrast, the Goldstone residue begins at order $k^2$, 
giving a low-energy peak in the dynamical structure factor at finite momentum. 
For diagonal flow ($Q=0$), however, the zero-momentum phase-channel transfer vanishes, 
leaving only higher-order nodal spectral leakage at finite momentum. 
The transfer thus tracks the $B_{1g}$ component of the phase gradient. 
Energy cuts in Fig. 3 (c) and 3(d) show enhanced cross-channel peaks 
for axial relative to diagonal flow.

\section*{Discussion}

We have developed a theoretical framework for a spin superfluid 
in an altermagnetic insulator~\cite{Qaiumzadeh2017,Zarzuela2021} with $d_{x^2-y^2}$-wave spin splitting~\cite{Smejkal2023Magnons,Syljuasen2025}, 
in which spin angular momentum is carried coherently by the phase of an easy-plane N\'{e}el order~\cite{Kim2016Thermal,Kim2016Quantum,Tserkovnyak2020}. 
Magnon-current coupling to N\'{e}el-vector dynamics induces rectification, 
rendering the system non-Hermitian and driving it into a nonequilibrium steady state.
The altermagnetic order reshapes the spin superfluid: 
an axial phase gradient hybridises the phase and out-of-plane modes, 
softens the Goldstone velocity, and transfers phase-channel spectral weight 
from the Goldstone mode to the gapped mode. At the shifted axial stability boundary, 
the Goldstone mode acquires a quadratic dispersion. 
These effects are absent for diagonal flow, establishing the superflow orientation 
as the control parameter of the altermagnetic fingerprint.
DFT calculations reveal the distinctive local crystal environment of $\mathrm{La_2O_3FeMnSe_2}$, 
in which the altermagnetic SSG generates chiral splitting of magnon bands and 
unconventional spin transport, supporting spin superfluidity. 
Strong external fields or elevated temperatures suppress the spin superfluid and restore
conventional magnon diffusion with exponential decay~\cite{NewRef17,NewRef05,NewRef18,NewRef06}
(Fig.~\ref{fig:fieldtemp}). These results motivate a full lattice spin-wave
treatment for material-specific predictions and underscore the potential of
altermagnets  as a versatile platform for low-dissipation spintronic devices 
that exploit superfluid spin transport\cite{Tserkovnyak2017,Liu2020Josephson}.


\section*{Methods}

See Supplementary Information.

\section*{Acknowledgements}

\begingroup\justifying\emergencystretch=6em
We thank Lu Xiao for helpful discussions. This work is supported by NKRDPC-2022YFA1402802, NSFC-92165204, Guangdong Provincial Key Laboratory of Magnetoelectric Physics and Devices (2026B1212170049), Research Center for Magnetoelectric Physics of Guangdong Province (2024B0303390001), Guangdong Provincial Quantum Science Strategic Initiative (GDZX2401010 and GDZX2601002), and Center for Advanced Study of SYSU.\par
\endgroup

\section*{Competing interests}

The authors declare no competing interests.

\section*{Author contributions}

Dao-Xin Yao designed the research. Dele Li performed the analytical
derivations and the Landau--Lifshitz--Gilbert simulations. Meng-Han Zhang
analysed the results and wrote the manuscript. All authors reviewed and approved
the final manuscript.

\bibliographystyle{group3}
\bibliography{references}

\end{document}


\maketitle

This Supplementary Information presents the continuum theory, collective modes and spectral response of the spin-superfluid state. All equation numbers refer to this Supplementary Information.
Coherent spin-superfluid transport in antiferromagnetic and ferromagnetic insulators has been described theoretically~\cite{Takei2014,Sonin2019}. Our analysis of the altermagnetic-insulator $\mathrm{La_2O_3FeMnSe_2}$ builds on spin-space-group studies of unconventional magnons in collinear magnets~\cite{Chen2025Magnons}.

\section*{\centering 1. From discrete exchange to the continuum theory}

The same-sublattice exchange along the two crystal axes, with bond length $a$, is
\begin{align}
 H_x={}&J_{2m}\bm S_A(\bm r)\cdot\bm S_A(\bm r+a\hat{\bm x})
       +J_{2n}\bm S_B(\bm r)\cdot\bm S_B(\bm r+a\hat{\bm x}),\nonumber\\
 H_y={}&J_{2n}\bm S_A(\bm r)\cdot\bm S_A(\bm r+a\hat{\bm y})
       +J_{2m}\bm S_B(\bm r)\cdot\bm S_B(\bm r+a\hat{\bm y}).
 \label{eq:discrete-exchange}
\end{align}
Here $A$ and $B$ label the magnetic sublattices, $J_{2m}$ and $J_{2n}$ are the same-sublattice exchanges as in the main text, and $S$ is the spin magnitude on each sublattice. To take the continuum limit, introduce the N\'eel and uniform-magnetisation fields,
\begin{equation}
 \frac{\bm{S}_A}{S}=\bn+\bmagn,\qquad
 \frac{\bm{S}_B}{S}=-\bn+\bmagn .
 \label{eq:fields}
\end{equation}

with $\bn\cdot\bmagn=0$ and $\bn^2+\bmagn^2=1$; for small $\bmagn$,
$\bn$ is treated as a unit vector to leading order. Writing
$\bn_\mu=\bn(\bm r+a\hat{\bm\mu})$ and
$\bmagn_\mu=\bmagn(\bm r+a\hat{\bm\mu})$, substitution into
Eq.~\eqref{eq:discrete-exchange} gives
\begin{align}
 H_x={}&S^2(J_{2m}+J_{2n})(\bn\cdot\bn_x+\bmagn\cdot\bmagn_x)\nonumber\\
 &+S^2\Delta J_0(\bn\cdot\bmagn_x+\bmagn\cdot\bn_x),
 \qquad \Delta J_0=J_{2m}-J_{2n}.
 \label{eq:exchange-decomposition}
\end{align}
The exchange-odd term changes sign along $y$. Expanding the fields,
\begin{align}
 \bn_\mu&=\bn+a\partial_\mu\bn
 +\frac{a^2}{2}\partial_\mu^2\bn+O(a^3),\nonumber\\
 \bmagn_\mu&=\bmagn+a\partial_\mu\bmagn
 +\frac{a^2}{2}\partial_\mu^2\bmagn+O(a^3).
 \label{eq:appendix-taylor-fields}
\end{align}

the constraint $\bn\cdot\bmagn=0$ removes the zeroth- and first-order terms and gives
\begin{equation}
 \bn\cdot\bmagn_\mu+\bmagn\cdot\bn_\mu
 =-a^2\partial_\mu\bn\cdot\partial_\mu\bmagn+O(a^3).
 \label{eq:appendix-cross-gradient}
\end{equation}

The exchange-odd continuum energy is therefore
\begin{equation}
 \mathcal H_{\rm alt}=
 -\mathcal C S^2a^2\Delta J_0
 \int\dd^2r\left(
 \partial_x\bn\cdot\partial_x\bmagn
 -\partial_y\bn\cdot\partial_y\bmagn\right),
 \label{eq:appendix-alt-energy-gradient}
\end{equation}

where $\mathcal C$ accounts for bond counting and unit-cell normalisation.
Integration by parts, with vanishing boundary terms, gives
\begin{align}
 &\int\dd^2r\left(
 \partial_x\bn\cdot\partial_x\bmagn
 -\partial_y\bn\cdot\partial_y\bmagn\right)\nonumber\\
 &\qquad=-\int\dd^2r\;\bmagn\cdot
 (\partial_x^2-\partial_y^2)\bn .
 \label{eq:appendix-integration-parts}
\end{align}

Absorbing the overall sign and normalisation into $\lambda$, the corresponding Lagrangian term is
\begin{align}
 \mathcal L_{\rm AM}&=\lambda\,\bmagn\cdot D\bn,
 \qquad D\equiv\partial_x^2-\partial_y^2,\nonumber\\
 \lambda&\propto S^2a^2(J_{2m}-J_{2n}).
 \label{eq:appendix-AM-invariant}
\end{align}

Together with the sublattice Berry term, the uniform-magnetisation cost,
the exchange stiffness $A_{\rm ex}$, and easy-plane anisotropy $K>0$, this gives
\begin{align}
 \mathcal L_{\bmagn}={}&s\bmagn\cdot(\bn\times\partial_t\bn)
 -\frac{\bmagn^2}{2\chi}+\lambda\bmagn\cdot D\bn \nonumber\\
 &-\frac{A_{\rm ex}}{2}(\nabla\bn)^2-\frac{K}{2}n_z^2 .
 \label{eq:Lm}
\end{align}
Here $K$ is the continuum easy-plane anisotropy coefficient, $s$ is the Berry-phase coefficient, and $\chi$ is the transverse susceptibility
associated with the uniform magnetisation.

Variation subject to $\bn\cdot\bmagn=0$ yields
\begin{equation}
 \bmagn=\chi P_\perp\left[s(\bn\times\partial_t\bn)+\lambda D\bn\right],
 \qquad P_\perp=\mathbf{1}-\bn\bn^{\mathsf T}.
 \label{eq:m-saddle}
\end{equation}

Substitution into Eq.~\eqref{eq:Lm} gives the N\'eel-field theory
\begin{align}
 \mathcal L_{\rm eff}={}&\frac{\rho}{2}(\partial_t\bn)^2
 +g_{\rm AM}(\bn\times\partial_t\bn)\cdot D\bn
 +\frac{g_{\rm AM}^2}{2\rho}\left|P_\perp D\bn\right|^2 \nonumber\\
 &-\frac{A_{\rm ex}}{2}(\nabla\bn)^2-\frac{K}{2}n_z^2,
 \label{eq:Leff}
\end{align}

Here $\rho=\chi s^2$ and $g_{\rm AM}=\chi s\lambda$.
The mixed term carries the $B_{1g}$ exchange anisotropy; the accompanying
$g_{\rm AM}^2/(2\rho)=\chi\lambda^2/2$ term follows from the same elimination of $\bmagn$.

\section*{\centering 2. Quadratic expansion about a spin-superfluid background}

For a state with fixed phase gradient $\bm q$, with $q=|\bm q|$, write
\begin{equation}
 \bn=\left(\sqrt{1-\theta^2}\cos\Phi,
 \sqrt{1-\theta^2}\sin\Phi,\theta\right),
 \qquad \Phi=\bm q\cdot\bm r+\varphi .
 \label{eq:superfluid-parametrisation}
\end{equation}

The background $\varphi=\theta=0$ is a finite-winding, current-carrying state.
Define $Q=q_x^2-q_y^2$ and $M=K-A_{\rm ex}q^2$. Since
$D\bn_0=-Q\bn_0$, one has $P_{\perp,0}D\bn_0=0$ and
$\bmagn_0=0$. Its energy density and spin current are
\begin{equation}
 \mathcal E_0=\frac{A_{\rm ex}}{2}q^2,\qquad j_i^z=-A_{\rm ex}q_i.
 \label{eq:background-current}
\end{equation}
With $\bar\Phi=\bm q\cdot\bm r$, introduce the local frame
\begin{align}
 \bar{\bm e}_\rho&=(\cos\bar\Phi,\sin\bar\Phi,0),\nonumber\\
 \bar{\bm e}_\varphi&=(-\sin\bar\Phi,\cos\bar\Phi,0),
 \qquad \bm e_z=(0,0,1).
 \label{eq:appendix-local-frame}
\end{align}

The linear fluctuation is $\delta\bn=\varphi\bar{\bm e}_\varphi+\theta\bm e_z$. Both the field and the projector must be varied:
\begin{align}
 \delta(P_\perp D\bn)
 &=P_{\perp,0}D\delta\bn+\delta P_\perp D\bn_0\nonumber\\
 &=(D\varphi)\bar{\bm e}_\varphi+(D\theta+Q\theta)\bm e_z,\nonumber\\
 |P_\perp D\bn|^2&=(D\varphi)^2+(D\theta+Q\theta)^2+O(\delta\bn^3).
 \label{eq:projector-expansion}
\end{align}
In particular, $\delta P_\perp D\bn_0$ cancels the $-Q\varphi$ contribution
and supplies $Q\theta\bm e_z$. Up to total derivatives, the mixed term becomes
\begin{align}
 (\bn\times\partial_t\bn)\cdot D\bn
 ={}&Q\theta\dot\varphi+\dot\varphi D\theta
 -\dot\theta D\varphi+O(\delta\bn^3)\nonumber\\
 ={}&Q\theta\dot\varphi+2\dot\varphi D\theta
 +O(\delta\bn^3).
 \label{eq:appendix-mixed-term-expansion}
\end{align}

The conventional terms, after dropping a constant and a boundary term, give
\begin{align}
 &\frac{\rho}{2}(\partial_t\bn)^2
 -\frac{A_{\rm ex}}{2}(\nabla\bn)^2-\frac{K}{2}n_z^2\nonumber\\
 &\quad=\frac{\rho}{2}(\dot\varphi^2+\dot\theta^2)
 -\frac{A_{\rm ex}}{2}\left[(\nabla\varphi)^2+(\nabla\theta)^2\right]
 -\frac{M}{2}\theta^2,
 \label{eq:appendix-sigma-expansion}
\end{align}

Combining these contributions yields
\begin{align}
 \mathcal L^{(2)}={}&\frac{\rho}{2}(\dot\varphi^2+\dot\theta^2)
 -\frac{A_{\rm ex}}{2}\left[(\nabla\varphi)^2+(\nabla\theta)^2\right]
 -\frac{M}{2}\theta^2 \nonumber\\
 &+g_{\rm AM}Q\theta\dot\varphi+2g_{\rm AM}\dot\varphi D\theta
 +\frac{g_{\rm AM}^2}{2\rho}\left[(D\varphi)^2+(D\theta+Q\theta)^2\right].
 \label{eq:L2}
\end{align}

For probe wave vector $\bk$ with $k=|\bk|$ and $e^{i(\bk\cdot\bm r-\omega t)}$, $D\to-\delta_{\bk}$ and
$D\theta+Q\theta\to(Q-\delta_{\bk})\theta$, where
\begin{equation}
 \delta_{\bk}=k_x^2-k_y^2,\qquad
 \Lambda_{\bk}=Q-2\delta_{\bk},\qquad
 \widetilde M=M-\frac{g_{\rm AM}^2Q^2}{\rho}.
 \label{eq:definitions}
\end{equation}
The quadratic action gives $\mathsf M(\omega,\bk)X=0$, with
$X=(\varphi,\theta)^{\mathsf T}$:
\begin{equation}
 \begin{pmatrix}
 \ell_{\varphi,\bk}-\rho\omega^2 & -i\omega g_{\rm AM}\Lambda_{\bk}\\
 i\omega g_{\rm AM}\Lambda_{\bk} & \ell_{\theta,\bk}-\rho\omega^2
 \end{pmatrix}
 \begin{pmatrix}\varphi\\ \theta\end{pmatrix}=0,
 \label{eq:mode-matrix}
\end{equation}

Here
\begin{align}
 \ell_{\varphi,\bk}&=A_{\rm ex}k^2-\frac{g_{\rm AM}^2}{\rho}\delta_{\bk}^2,\nonumber\\
 \ell_{\theta,\bk}&=A_{\rm ex}k^2+M-\frac{g_{\rm AM}^2}{\rho}(Q-\delta_{\bk})^2.
 \label{eq:diagonal-kernels}
\end{align}

The two branches follow from $\det\mathsf M=0$:
\begin{equation}
 \omega_\pm^2(\bk)=\frac{\mathcal B_{\bk}\pm
 \sqrt{\mathcal B_{\bk}^2-4\rho^2
 \ell_{\varphi,\bk}\ell_{\theta,\bk}}}{2\rho^2},
 \label{eq:mode-spectrum}
\end{equation}

where
\begin{equation}
 \mathcal B_{\bk}=\rho(\ell_{\varphi,\bk}+\ell_{\theta,\bk})
 +g_{\rm AM}^2\Lambda_{\bk}^2 .
 \label{eq:mode-trace}
\end{equation}

At zero momentum, $\ell_{\varphi,0}=0$ and $\ell_{\theta,0}=\widetilde M$, so
\begin{equation}
 \det\mathsf M(\omega,\bm 0)
 =\rho\omega^2\left(\rho\omega^2-M\right).
 \label{eq:appendix-zero-momentum-determinant}
\end{equation}

Thus $\omega_-(0)=0$ and $\omega_+^2(0)=M/\rho$: the static $g_{\rm AM}^2Q^2$ correction
cancels the dynamic mixing contribution in the uniform gap.
For $Q\ne0$, expansion of Eq.~\eqref{eq:mode-spectrum} gives
\begin{align}
 \omega_-^2(\bk)={}&
 \frac{A_{\rm ex}\widetilde M}{\rho M}k^2+O(k^4),
 \label{eq:sound-velocity}\\
 \omega_+^2(\bk)={}&
 \frac{M}{\rho}+
 \left(\frac{A_{\rm ex}}{\rho}+
 \frac{A_{\rm ex}g_{\rm AM}^2Q^2}{\rho^2M}\right)k^2
 -\frac{2g_{\rm AM}^2Q}{\rho^2}\delta_{\bk}+O(k^4).
 \label{eq:gapped-mode-Qnonzero}
\end{align}

The Goldstone velocity is reduced by the background-flow component $Q$,
but is independent of the propagation direction to leading order.
The upper branch instead has a $k^2$ anisotropy proportional to
$Q\delta_{\bk}=Qk^2\cos(2\phi_{\bk})$, where
$\bk=k(\cos\phi_{\bk},\sin\phi_{\bk})$.
For zero or diagonal flow ($Q=0$) at fixed $M>0$,
\begin{align}
 \omega_-^2={}&\frac{A_{\rm ex}}{\rho}k^2-\frac{g_{\rm AM}^2}{\rho^2}\delta_{\bk}^2
 -\frac{4A_{\rm ex}g_{\rm AM}^2}{\rho^2M}k^2\delta_{\bk}^2+O(k^8),
 \label{eq:Q0-gapless}\\
 \omega_+^2={}&\frac{M}{\rho}+\frac{A_{\rm ex}}{\rho}k^2
 +\frac{3g_{\rm AM}^2}{\rho^2}\delta_{\bk}^2
 +\frac{4A_{\rm ex}g_{\rm AM}^2}{\rho^2M}k^2\delta_{\bk}^2+O(k^8).
 \label{eq:Q0-spectrum}
\end{align}

Since $\delta_{\bk}^2=k^4\cos^2(2\phi_{\bk})$, these corrections are maximal
along the crystal axes and vanish along the diagonals. The leading Goldstone
velocity remains $\sqrt{A_{\rm ex}/\rho}$, as for $g_{\rm AM}=0$ .

The long-wavelength stability condition is
\begin{equation}
 \widetilde M=K-A_{\rm ex}q^2-\frac{g_{\rm AM}^2}{\rho}(q_x^2-q_y^2)^2>0.
 \label{eq:stability}
\end{equation}
It reduces the axial critical gradient, while $Q=0$ or $g_{\rm AM}=0$ gives
$q_c=\sqrt{K/A_{\rm ex}}$. At $\widetilde M=0$ with fixed $g_{\rm AM}Q\ne0$,
\begin{equation}
 \omega_-^2(\bk)=\frac{A_{\rm ex}}{g_{\rm AM}^2Q^2}
 \left[A_{\rm ex}+\frac{2g_{\rm AM}^2Q}{\rho}\cos(2\phi_{\bk})\right]k^4
 +O(k^6).
 \label{eq:critical-mode}
\end{equation}

The Goldstone mode is therefore quadratic when the $k^4$ coefficient is positive. Stability checks must also account for the direction dependence of this coefficient. For diagonal flow ($Q=0$) or $g_{\rm AM}=0$, the boundary instead occurs at $M\to0^+$: the Goldstone mode remains linear and the out-of-plane mode softens at $\bk=\bm{0}$. The $M\to0^+$ limit follows directly from Eq.~\eqref{eq:mode-spectrum}. These limits are distinct, and Eq.~\eqref{eq:critical-mode} applies only for $g_{\rm AM}Q\ne0$.
These results establish that the low-energy collective modes depend on both the direction and magnitude of the spin superflow. To resolve the behaviour near the stability boundary, we examine the Goldstone dispersion as the phase gradient approaches its critical value.

\begin{figure}[!htbp]
 \centering
 \refstepcounter{figure}

 \SupplementIncludegraphics[width=\textwidth]{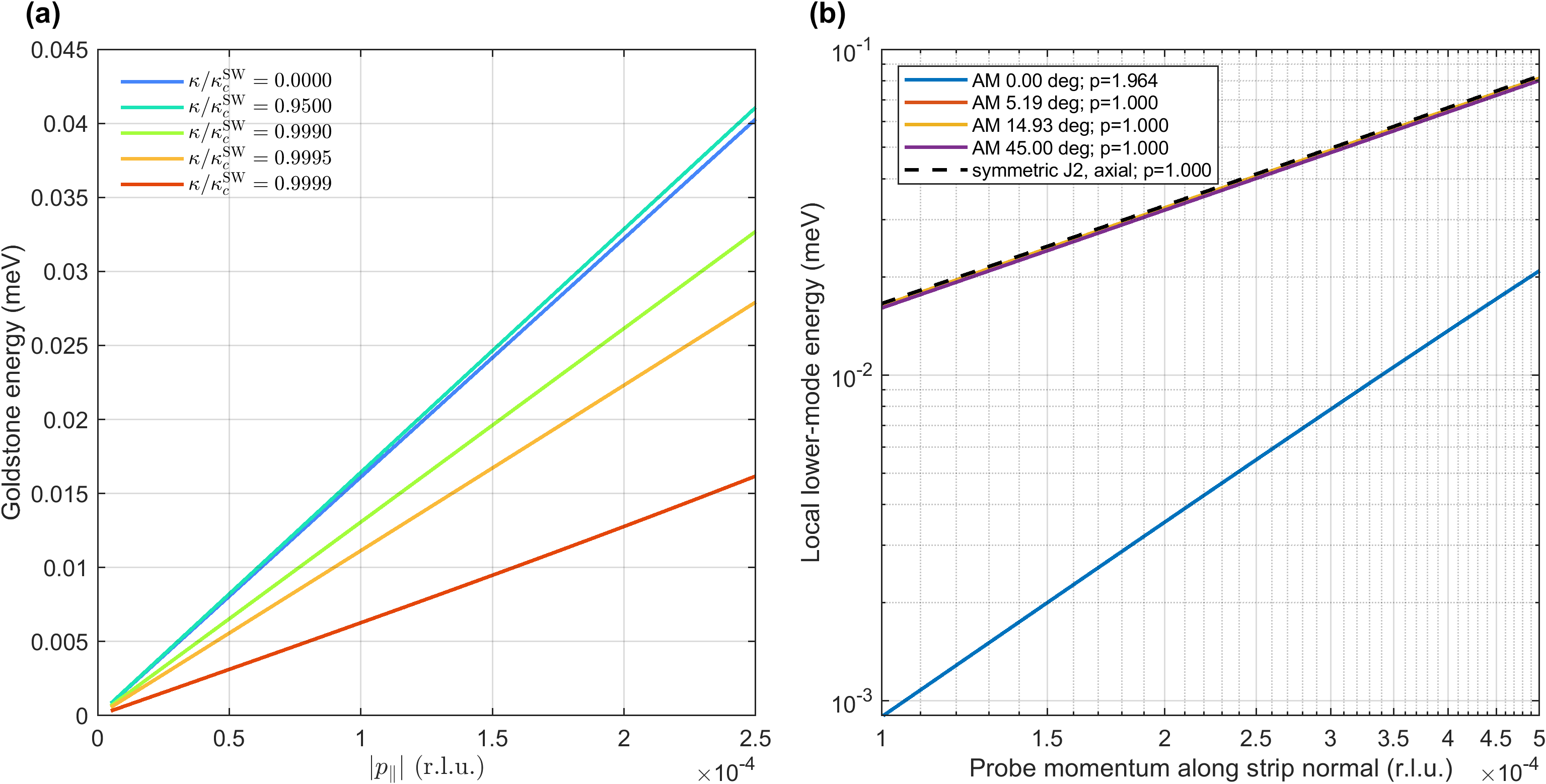}
 \par\begin{minipage}{\textwidth}\raggedright\small\textbf{Supplementary Figure \thefigure.} Goldstone-mode softening near the critical spin flow. \textbf{(a)} Goldstone-mode energy as a function of the probe momentum $|p_{\parallel}|$ for several values of $\kappa/\kappa_c^{\mathrm{SW}}$. The mode softens as the axial flow approaches the critical value. \textbf{(b)} Low-energy lower-mode dispersion for different altermagnetic transport directions and the symmetric-$J_2$ axial reference. Here $p_{\parallel}$ denotes the component of the probe wave vector along the transport direction, and $\kappa/\kappa_c^{\mathrm{SW}}$ is the phase gradient normalised by its spin-wave critical value. The symmetric-$J_2$ label denotes the equal-exchange reference. The exponents shown in the legend are obtained from power-law fits over the displayed momentum window.
 \label{fig:supp-critical}
 \end{minipage}
\end{figure}

\begin{figure}[!htbp]
 \centering
 \refstepcounter{figure}

 \SupplementIncludegraphics[width=0.80\textwidth]{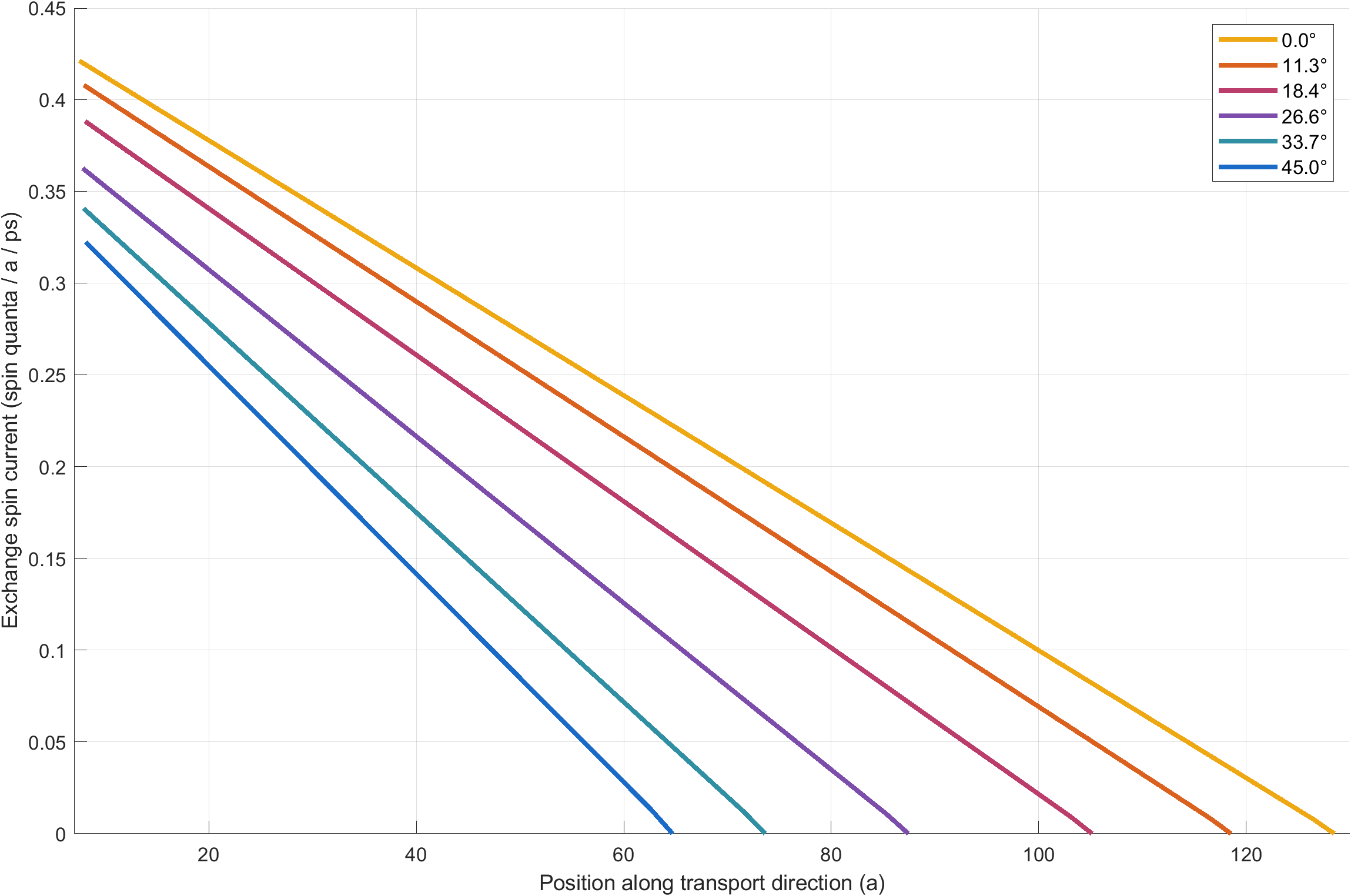}
 \par\begin{minipage}{\textwidth}\raggedright\small\textbf{Supplementary Figure \thefigure.} Direction-dependent exchange spin-current profiles. Exchange spin current as a function of position along the transport direction for the angles indicated in the legend. The position is expressed in units of the lattice constant $a$.
 \label{fig:supp-directions}
 \end{minipage}
\end{figure}

Supplementary Fig.~\ref{fig:supp-critical}a shows the low-energy Goldstone mode softening as the axial flow approaches the critical spin-wave flow $\kappa_c^{\mathrm{SW}}$; the softening is strongest near the critical flow, confirming that the low-energy response is sensitive to the phase gradient. Supplementary Fig.~\ref{fig:supp-critical}b shows the directional dependence: within the displayed momentum window, the axial altermagnetic case follows a power law with fitted exponent close to two, whereas the non-axial cases and the symmetric-exchange axial reference are nearly linear. This agrees with the analytical distinction between the axial critical mode and the non-axial or conventional reference cases. To examine the spatial dependence of spin transport, we plot the exchange spin current versus position along the transport direction for angles from $0^\circ$ to $45^\circ$ (Supplementary Fig.~\ref{fig:supp-directions}).

The current profiles are direction-dependent: in the cases shown, the axial profile carries a larger current and extends farther than the diagonal profile. This contrast holds for the device and boundary conditions considered. The spectral response of magnetic excitations in altermagnets has been studied, including its connection to magnon quantum geometry~\cite{Syljuasen2025}. Polarization-resolved Raman scattering has also probed amplitude and magnon modes in the antiferromagnet $\mathrm{Ca_2RuO_4}$~\cite{Souliou2017}. In the continuum description, the local channels $\varphi$ and $\theta=\delta n_z$ represent in-plane and out-of-plane N\'{e}el fluctuations. We introduce sources $\mathcal L_{\rm src}=f_\varphi\varphi+f_\theta\theta$ to compute the corresponding retarded susceptibilities, and set $z=\omega+i0^+$. With the convention $\boldsymbol\chi_X^R=-\mathsf M^{-1}$,

\begin{align}
 \boldsymbol{\chi}^R_X(z,\bk)={}&
 -\frac{1}{\Delta(z,\bk)}
 \begin{pmatrix}
 \ell_{\theta,\bk}-\rho z^2 & izg_{\rm AM}\Lambda_{\bk}\\
 -izg_{\rm AM}\Lambda_{\bk} & \ell_{\varphi,\bk}-\rho z^2
 \end{pmatrix},\nonumber\\
 X={}&
 \begin{pmatrix}\varphi\\ \theta\end{pmatrix},
 \label{eq:retarded-susceptibility}
\end{align}

where
\begin{align}
 \Delta(z,\bk)={}&
 (\ell_{\varphi,\bk}-\rho z^2)
 (\ell_{\theta,\bk}-\rho z^2)
 -g_{\rm AM}^2z^2\Lambda_{\bk}^2 \nonumber\\
 ={}&\rho^2(z^2-\omega_-^2)(z^2-\omega_+^2).
 \label{eq:response-determinant}
\end{align}

This convention differs by a minus sign from the algebraic source response
$\mathsf M^{-1}$ and gives nonnegative diagonal spectral densities at positive frequencies,
\begin{equation}
 \mathcal S_a(\omega,\bk)
 =-\frac{1}{\pi}\operatorname{Im}\chi^R_{aa}(\omega,\bk),
 \qquad a=\varphi,\theta.
 \label{eq:spectral-density-definition}
\end{equation}

The off-diagonal entries describe phase-sensitive mixed responses. For each diagonal channel, write
\begin{equation}
 \chi^R_{aa}(z,\bk)=
 \frac{\mathcal R_-^a(\bk)}{z^2-\omega_-^2(\bk)}
 +\frac{\mathcal R_+^a(\bk)}{z^2-\omega_+^2(\bk)}.
 \label{eq:residue-decomposition}
\end{equation}

With
\begin{equation}
 \mathcal D_{\bk}=
 \sqrt{\mathcal B_{\bk}^2-4\rho^2
 \ell_{\varphi,\bk}\ell_{\theta,\bk}},
 \qquad
 \Xi_{\bk}=\ell_{\theta,\bk}-\ell_{\varphi,\bk},
 \label{eq:residue-discriminant}
\end{equation}

the residues are
\begin{align}
 \mathcal R_-^\theta&=
 \frac{\mathcal D_{\bk}-\rho\Xi_{\bk}-g_{\rm AM}^2\Lambda_{\bk}^2}
 {2\rho\mathcal D_{\bk}},
 &\mathcal R_+^\theta&=
 \frac{\mathcal D_{\bk}+\rho\Xi_{\bk}+g_{\rm AM}^2\Lambda_{\bk}^2}
 {2\rho\mathcal D_{\bk}}, \label{eq:theta-residues}\\
 \mathcal R_-^\varphi&=
 \frac{\mathcal D_{\bk}+\rho\Xi_{\bk}-g_{\rm AM}^2\Lambda_{\bk}^2}
 {2\rho\mathcal D_{\bk}},
 &\mathcal R_+^\varphi&=
 \frac{\mathcal D_{\bk}-\rho\Xi_{\bk}+g_{\rm AM}^2\Lambda_{\bk}^2}
 {2\rho\mathcal D_{\bk}}. \label{eq:phi-residues}
\end{align}

The positive-frequency peak area is
\begin{equation}
 \mathcal Z_\nu^a(\bk)=
 \frac{\mathcal R_\nu^a(\bk)}{2\omega_\nu(\bk)}.
 \label{eq:oscillator-strength}
\end{equation}

The unbroadened spectrum is
\begin{equation}
 \mathcal S_a(\omega>0,\bk)=
 \sum_{\nu=\pm}\mathcal Z_\nu^a(\bk)
 \delta[\omega-\omega_\nu(\bk)].
 \label{eq:delta-spectral-function}
\end{equation}

A common phenomenological linewidth $\eta>0$, implemented by $z=\omega+i\eta$, gives
\begin{equation}
 \mathcal S_a(\omega,\bk)=
 \sum_{\nu=\pm}\frac{\mathcal R_\nu^a(\bk)}{\pi}
 \frac{2\omega\eta}
 {[\omega^2-\omega_\nu^2(\bk)-\eta^2]^2+4\omega^2\eta^2}.
 \label{eq:broadened-spectral-function}
\end{equation}

At $g_{\rm AM}=0$, the two modes obey the selection rule
\begin{equation}
 g_{\rm AM}=0:\qquad
 \mathcal R_-^\varphi=\mathcal R_+^\theta=\frac{1}{\rho},
 \qquad
 \mathcal R_-^\theta=\mathcal R_+^\varphi=0.
 \label{eq:afm-selection-rule}
\end{equation}

For $Q=0$, only the phase channel carries the Goldstone mode and only the out-of-plane channel carries the gapped mode. For $Q\ne0$ in the stable regime, the zero-momentum residues are
\begin{align}
 \mathcal R_-^\varphi(\bm{0})&=\frac{\widetilde M}{\rho M},
 &\mathcal R_+^\varphi(\bm{0})&=\frac{g_{\rm AM}^2Q^2}{\rho^2M},
 \label{eq:Qnonzero-phase-residues}\\
 \mathcal R_-^\theta(\bm{0})&=0,
 &\mathcal R_+^\theta(\bm{0})&=\frac{1}{\rho}.
 \label{eq:Qnonzero-theta-residues}
\end{align}

As $\widetilde M\to0^+$, phase-channel weight shifts from the Goldstone to the gapped mode, whereas the out-of-plane Goldstone residue remains zero at $k=0$. For axial flow, the transferred fraction is $g_{\rm AM}^2q^4/(\rho K)+O(q^6)$; diagonal flow shows no transfer at $k=0$. We quote residues rather than $\mathcal Z=\mathcal R/(2\omega)$, which diverges for the zero-frequency Goldstone mode.

At finite small momentum and fixed $\widetilde M>0$,
Eq.~\eqref{eq:sound-velocity} and the residues give
\begin{equation}
 \mathcal Z_-^\theta(\bk)\propto|\bk|\qquad(Q\ne0).
 \label{eq:theta-oscillator-selection}
\end{equation}
Axial flow therefore activates an out-of-plane Goldstone peak at finite
momentum and a gapped phase peak with finite $\mathcal Z_+^\varphi(0)$.
For $Q=0$ at fixed $M>0$, the otherwise dark channels instead have
\begin{align}
 \mathcal R_-^{\theta}(\bk)&=
 \frac{4A_{\rm ex}g_{\rm AM}^2}{\rho^2M^2}k^2\delta_{\bk}^2+O(k^8),\\
 \mathcal R_+^{\varphi}(\bk)&=
 \frac{4g_{\rm AM}^2}{\rho^2M}\delta_{\bk}^2+O(k^6).
 \label{eq:Qzero-dark-residues}
\end{align}

Hence,
\begin{equation}
 \mathcal Z_-^\theta=O(k^5),\qquad
 \mathcal Z_+^\varphi=O(k^4).
 \label{eq:Qzero-oscillator-strengths}
\end{equation}

Both weights vanish on $k_x=\pm k_y$. Finite $Q$ therefore produces a much
stronger long-wavelength transfer .

In each channel the residues satisfy
\begin{equation}
 \mathcal R_-^\varphi+\mathcal R_+^\varphi
 =\mathcal R_-^\theta+\mathcal R_+^\theta
 =\frac{1}{\rho}.
 \label{eq:residue-sum-rule}
\end{equation}

Equivalently,
\begin{equation}
 \int_0^\infty \dd\omega\;2\omega\,
 \mathcal S_a(\omega,\bk)=\frac{1}{\rho},
 \qquad a=\varphi,\theta.
 \label{eq:spectral-sum-rule}
\end{equation}

The unweighted peak area in the zero-linewidth limit is instead
\begin{equation}
 \int_0^\infty \dd\omega\;\mathcal S_a(\omega,\bk)=
 \sum_{\nu=\pm}\frac{\mathcal R_\nu^a(\bk)}
 {2\omega_\nu(\bk)},
 \label{eq:unweighted-spectral-weight}
\end{equation}

and depends on momentum. The coupling redistributes spectral weight while preserving the frequency-weighted sum.

\section*{\centering 3. First-order formulation retaining the uniform magnetisation}

For $\bm q=0$, the same modes can be obtained without first eliminating
$\bmagn$. Here $\theta$ denotes the tilt angle, which coincides with the
out-of-plane fluctuation in Eq.~\eqref{eq:superfluid-parametrisation} to linear order.
Use the angular parametrisation
\begin{equation}
 \bn=(\cos\theta\cos\varphi,\cos\theta\sin\varphi,\sin\theta),
 \label{eq:appendix-D-n-parametrisation}
\end{equation}

and tangent basis
\begin{align}
 \bm e_\theta={}&(-\sin\theta\cos\varphi,
 -\sin\theta\sin\varphi,\cos\theta),\nonumber\\
 \bm e_\varphi={}&(-\sin\varphi,\cos\varphi,0).
 \label{eq:appendix-D-tangent-frame}
\end{align}

Writing $\bmagn=m_\theta\bm e_\theta+m_\varphi\bm e_\varphi$ enforces
$\bn\cdot\bmagn=0$. The Berry term is
\begin{equation}
 \bmagn\cdot(\bn\times\dot\bn)
 =m_\theta\cos\theta\,\dot\varphi-m_\varphi\dot\theta.
 \label{eq:appendix-D-Berry}
\end{equation}

Expansion of Eq.~\eqref{eq:Lm} about $\varphi=\theta=0$ gives
\begin{align}
 \mathcal L^{(2)}_{\bmagn}={}&
 s\left(m_\theta\dot\varphi-m_\varphi\dot\theta\right)
 -\frac{m_\varphi^2+m_\theta^2}{2\chi}\nonumber\\
 &+\lambda\left(m_\varphi D\varphi+m_\theta D\theta\right)\nonumber\\
 &-\frac{A_{\rm ex}}{2}\left[(\nabla\varphi)^2+(\nabla\theta)^2\right]
 -\frac{K}{2}\theta^2.
 \label{eq:appendix-D-quadratic-Lagrangian}
\end{align}

The Euler--Lagrange equations are
\begin{align}
 0={}&-s\dot\theta-\frac{m_\varphi}{\chi}+\lambda D\varphi,
 \label{eq:appendix-D-EL-mphi}\\
 0={}&s\dot\varphi-\frac{m_\theta}{\chi}+\lambda D\theta,
 \label{eq:appendix-D-EL-mtheta}\\
 0={}&-s\dot m_\theta+\lambda Dm_\varphi+A_{\rm ex}\nabla^2\varphi,
 \label{eq:appendix-D-EL-phi}\\
 0={}&s\dot m_\varphi+\lambda Dm_\theta
 +A_{\rm ex}\nabla^2\theta-K\theta.
 \label{eq:appendix-D-EL-theta}
\end{align}

The first two give
\begin{equation}
 m_\varphi=\chi\left(-s\dot\theta+\lambda D\varphi\right),
 \qquad
 m_\theta=\chi\left(s\dot\varphi+\lambda D\theta\right).
 \label{eq:appendix-D-m-saddle}
\end{equation}

For the field ordering $\Psi_{\bk}=(m_\varphi,m_\theta,\varphi,\theta)^{\mathsf T}$, the Fourier equations are
\begin{align}
 \mathsf K(\omega,\bk)\Psi_{\bk}&=0,\nonumber\\
 \mathsf K(\omega,\bk)&=
 \begin{pmatrix}
 \chi^{-1} & 0 & \lambda\delta_{\bk} & -is\omega\\
 0 & \chi^{-1} & is\omega & \lambda\delta_{\bk}\\
 \lambda\delta_{\bk} & -is\omega & A_{\rm ex}k^2 & 0\\
 is\omega & \lambda\delta_{\bk} & 0 & A_{\rm ex}k^2+K
 \end{pmatrix}.
 \label{eq:appendix-D-four-by-four}
\end{align}

Eliminating the upper-left block $\chi^{-1}\mathbf{1}$ gives the Schur complement
\begin{align}
 \mathsf M_0(\omega,\bk)={}&
 \begin{pmatrix}
 \ell_{\bk}-\rho\omega^2 & 2i\omega g_{\rm AM}\delta_{\bk}\\
 -2i\omega g_{\rm AM}\delta_{\bk} & \ell_{\bk}+K-\rho\omega^2
 \end{pmatrix},\nonumber\\
 \ell_{\bk}={}&A_{\rm ex}k^2-\frac{g_{\rm AM}^2}{\rho}\delta_{\bk}^2,
 \label{eq:appendix-D-Schur-matrix}
\end{align}

This is Eq.~\eqref{eq:mode-matrix} at $\bm q=0$, where $M=K$ and
$\Lambda_{\bk}=-2\delta_{\bk}$. Correspondingly,
\begin{equation}
 \det\mathsf K=\chi^{-2}
 \left[(\ell_{\bk}-\rho\omega^2)
 (\ell_{\bk}+K-\rho\omega^2)
 -4g_{\rm AM}^2\omega^2\delta_{\bk}^2\right].
 \label{eq:appendix-D-determinant}
\end{equation}

\begin{figure}[!htbp]
 \centering
 \refstepcounter{figure}

 \SupplementIncludegraphics[width=0.80\textwidth]{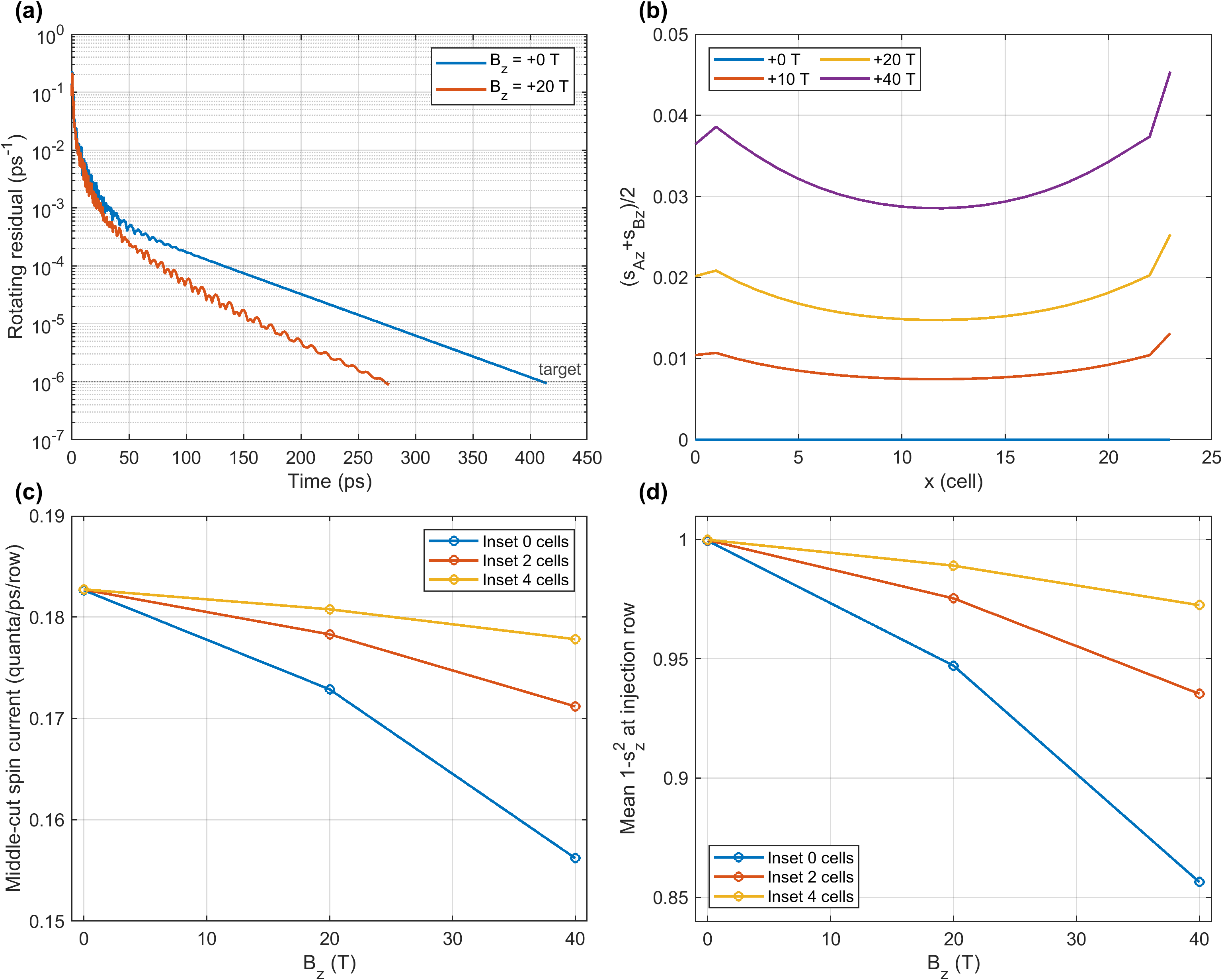}
 \par\begin{minipage}{\textwidth}\raggedright\small\textbf{Supplementary Figure \thefigure.} Magnetic-field response and effects of the contact position. \textbf{(a)} Rotating-frame residual as a function of time for $B_z=0$ and $20$ T. The horizontal line marks the target residual. \textbf{(b)} Spatial profiles of $(s_{Az}+s_{Bz})/2$ for the fields indicated in the legend. \textbf{(c)} Middle-cut exchange spin current as a function of $B_z$, comparing contacts moved inward by 0, 2 and 4 cells. \textbf{(d)} Corresponding mean value of $1-s_z^2$ at the injection row.
 \label{fig:supp-field}
 \end{minipage}
\end{figure}

The spectrum therefore comprises two positive-frequency branches, obtained from Eq.~\eqref{eq:mode-spectrum} by setting $\ell_{\varphi,\bk}=\ell_{\bk}$, $\ell_{\theta,\bk}=\ell_{\bk}+K$, and $\mathcal B_{\bk}=\rho(2\ell_{\bk}+K)+4g_{\rm AM}^2\delta_{\bk}^2$. The magnetisation components are auxiliary: retaining them introduces no additional spectral pole and leaves the N\'{e}el-channel residues unchanged. We examine how an out-of-plane field and contact position affect the response. Supplementary Fig.~\ref{fig:supp-field}a shows the rotating-frame residual versus time for $B_z=0$ and $20$ T. In both cases, the residual decays toward the target value, numerically confirming relaxation to a steady state.

Supplementary Fig.~\ref{fig:supp-field}b shows the spatial profile of the mean out-of-plane spin component $(s_{Az}+s_{Bz})/2$ for different magnetic fields. A finite field induces a nonzero out-of-plane component whose amplitude and spatial profile depend on the field strength. Supplementary Fig.~\ref{fig:supp-field}c,d shows the effect of moving the contacts inward. As the field increases, both the spin current at the middle cut and the mean $1-s_z^2$ at the injection row decrease in the cases shown. Moving the contacts inward suppresses these changes for the geometries considered. These results show that both the field and contact geometry control the calculated spin transport.

\section*{References}
\bibliography{supp_references}